\documentclass[twocolumn,british,english,aps,prb,twocolumns]{revtex4}
\usepackage[T1]{fontenc}
\usepackage[latin9]{inputenc}
\usepackage{color}
\usepackage{graphicx}
\usepackage{esint}

\makeatletter

\providecommand{\tabularnewline}{\\}

\@ifundefined{textcolor}{}
{%
 \definecolor{BLACK}{gray}{0}
 \definecolor{WHITE}{gray}{1}
 \definecolor{RED}{rgb}{1,0,0}
 \definecolor{GREEN}{rgb}{0,1,0}
 \definecolor{BLUE}{rgb}{0,0,1}
 \definecolor{CYAN}{cmyk}{1,0,0,0}
 \definecolor{MAGENTA}{cmyk}{0,1,0,0}
 \definecolor{YELLOW}{cmyk}{0,0,1,0}
}

\makeatother

\usepackage{babel}
\begin{document}

\title{Large scale GW-BSE calculations with N$^{3}$ scaling: excitonic
effects in dye sensitised solar cells}

\author{Margherita Marsili}

\affiliation{Dipartimento di Fisica e Astronomia, Università di Padova, via Marzolo
8, I-35131 Padova, Italy }

\author{Edoardo Mosconi}

\affiliation{Istituto CNR di Scienze e Tecnologie Molecolare, via Elce di Sotto
8, I-06123 Perugia, Italy}

\affiliation{CompuNet, Istituto Italiano di Tecnologia, Via Morego 30, 16163 Genova,
Italy.}

\author{Filippo De Angelis}

\affiliation{Istituto CNR di Scienze e Tecnologie Molecolare, via Elce di Sotto
8, I-06123 Perugia, Italy}

\affiliation{CompuNet, Istituto Italiano di Tecnologia, Via Morego 30, 16163 Genova,
Italy.}

\author{Paolo Umari}

\affiliation{Dipartimento di Fisica e Astronomia, Università di Padova, via Marzolo
8, I-35131 Padova, Italy}
\begin{abstract}
Excitonic effects due to electron-hole coupling play a fundamental
role in renormalising energy levels in dye sensitised and organic
solar cells determining the driving force for electron extraction.
We show that first-principles calculations based on many-body perturbation
theory within the GW-BSE approach provide a quantitative picture of
interfacial excited state energetics in organic dye-sensitized TiO$_{2}$,
delivering a general rule for evaluating relevant energy levels.To
perform GW-BSE calculations in such large systems we introduce a new
scheme based on maximally localized Wannier\textquoteright s functions.
With this method the overall scaling of GW-BSE calculations is reduced
from O(N$^{4}$) to O(N$^{3}$).
\end{abstract}
\maketitle

\section{Introduction}

Excitonic effects due to electron-hole coupling play a fundamental
role in renormalising energy levels in dye sensitised and organic
solar cells determining the driving force for electron extraction.
In particular, dye-sensitised solar cells (DSSCs) \cite{Oregan91}
are one of the most promising technologies for clean energy production.
Although \foreignlanguage{british}{currently} commercialized only
for niche applications, they are important as prototypical cells sharing
several features with organic and perovskite solar cells\cite{Lee12,Burschka13}. 

\begin{figure}
\includegraphics[scale=0.22]{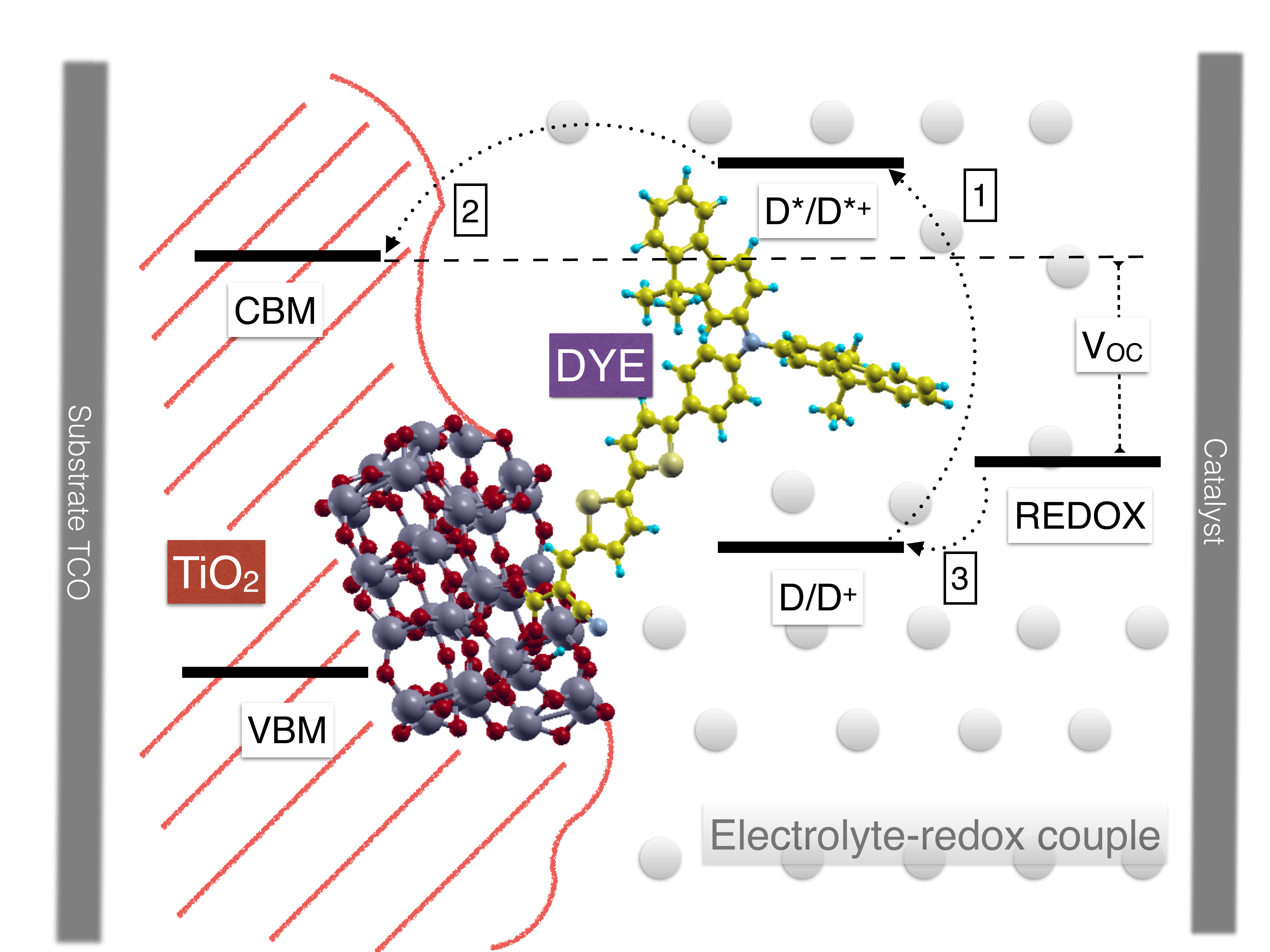}

\protect\caption{\label{fig:1}Schematic of a DSSC. The balls and sticks graph illustrates
the atomic structure of the model TiO$_{2}$ cluster sensitised with
the organic JK2 dye studied here. The main energy levels are: (D/D$^{+}$)
dye ground state ionisation potential, (D$*$/D{*}$^{+}$) dye excited
state ionisation potential, (CBM) TiO$_{2}$ conduction band minimum,
(VBM) TiO$_{2}$ valence band maximum, (REDOX) level of the redox
couple. $V_{\mathrm{oc}}$ is the open circuit voltage of the cell.}
 
\end{figure}

In DSSCs a semi-conducting oxide (typically meso or nano structured
TiO$_{2}$ ) supports a layer of adsorbed dye molecules. The oxide
is in contact with a transparent conductive oxide (TCO) and the dye
molecules are in contact with a hole transporting medium which can
be a polymeric solid or a redox couple in a liquid electrolyte (see
Fig. \ref{fig:1}). When a photon is adsorbed by the dye, one of its
valence electrons is promoted to an empty level. Provided that its
energy is higher than the TiO$_{2}$ conduction band minimum (CBM),
it can be injected into the TiO$_{2}$ and travel towards the electrode
leaving a hole in the dye. This can be eventually neutralised provided
that its energy level is lower than that of the redox couple. Therefore
the functioning of such a device crucially relies on the alignment
among the energy levels of its components (Fig. \ref{fig:1}). These
refer to both charged excitations (as TiO$_{2}$ band edges) and neutral
excitations (as the dye optical gap). 

Density functional theory (DFT) is nowadays the standard approach
for the modelling of complex systems as DSSCs providing an accurate
description of the atomistic structure. Unfortunately, energy levels
can be only approximately reproduced by DFT due to issues related
to the well-known band-gap underestimation \cite{De Angelis 14}. 

Many-body perturbation theory approaches (MBPT) can overcome these
issues\cite{Onida02}. Recently, the GW method \cite{Hedin65,Hybertsen86},
in which the electronic self-energy is expressed as the product of
the one-particle Green's function $G$ with the screened Coulomb interaction
$W$, has been applied to models of DSSCs leading to the successful
prediction of open-circuit voltages\cite{Umari13,Verdi14}. However,
the GW approach is appropriate only for the evaluation of charged
excitations as it does not take into account electron-hole interactions.
This is a major drawback for the calculation of the energy levels
of the excited dye and even more severe for those of dye-sensitised
semiconductors. 

Electron-hole interactions are introduced into the first-principle
MBPT framework through the Bethe Salpeter equation (BSE) approach
which gives access to excitonic energies and amplitudes \cite{Strinati88,Rohlfing00}.\textcolor{black}{{}
However, the high computational cost of BSE approaches and in particular
the $N^{4}$ scaling hinders their application to large model structures
as those required for treating DSSCs.}\textcolor{red}{{} }\textcolor{black}{Methods
presenting better $N^{3}$ scaling have been recently presented, however
they are limited either to finite systems \cite{Ljungberg15} or to
a stochastic sampling of the response functions\cite{Rabani15}.}

Here, we introduce a general BSE method which allows much faster calculations
on large systems thanks to a favorable $N^{3}$ scaling with respect
to the system size. Indeed, we express the excitonic amplitudes as
a set of excitations in correspondence with the maximally localised
Wannier's functions \cite{Marzari97} which define the manifold of
occupied states. In this way, we can avoid the calculation in the
BSE equation of terms related to non-overlapping Wannier's functions,
avoiding at the same time any explicit sum over empty one-particle
DFT orbitals\cite{Rocca10}. First, we validate our method addressing
the relatively small C$_{60}$ molecule. Then, we apply it to a large
model of a DSSC comprising the all-organic JK2 \cite{Kim06} dye adsorbed
on a TiO$_{2}$ cluster.

\section{Method}

The MBPT formalism gives a BSE in which a Hamiltonian-like operator
$\hat{H}^{eh}$ accounts for the electron-hole interactions\cite{Rohlfing00}.
Excitation energies $\Omega_{S}$ and amplitudes $\Theta$ are obtained
solving:

\begin{equation}
\hat{H}^{eh}\left|\Theta\right\rangle =\Omega_{S}\left|\Theta\right\rangle 
\end{equation}
Within the Tamm-Dancoff approximation the many-body excitonic (excited)
state $\left|\Theta\right\rangle $ is built starting from the ground
state $\left|\Phi_{0}\right\rangle $ which is taken to be a single
Slater determinant:
\begin{equation}
\left|\Phi_{0}\right\rangle =\frac{1}{\sqrt{N_{v}!}}\det\left|\psi_{1}\ldots\psi_{N_{v}}\right|
\end{equation}
where $N_{v}$ is the number of valence states. For simplicity we
omit here spin indices. Spin multiplicity will be considered later.
The one-body states $\psi_{i}$ are the eigenstate of a Kohn-Sham
Hamiltonian:

\begin{equation}
\epsilon_{i}^{KS}\left|\psi_{i}\right\rangle =H_{KS}\left|\psi_{i}\right\rangle 
\end{equation}
The GW method in its diagonal approximation is then used to obtain
for each $\psi_{i}$ the corresponding quasi-particle energy $\epsilon_{i}$.
We write, using the second quantization framework:

\begin{equation}
\left|\Phi_{0}\right\rangle =\frac{1}{\sqrt{N_{v}!}}\left(\prod_{v}\hat{a}_{v}^{\dagger}\right)\left|0\right\rangle 
\end{equation}
and

\begin{equation}
\left|\Theta\right\rangle =\sum_{vc}A_{vc}\hat{a}_{v}\hat{a}_{c}^{\dagger}\left|\Phi_{0}\right\rangle 
\end{equation}
 where $\hat{a}_{i}$ and $\hat{a}_{i}^{\dagger}$ are the annihilation
and construction operators for the $i$-th state, and the $v$ and
$c$ indices run over the occupied and empty states, respectively.
We define\cite{Rocca10,Walker06}:

\begin{equation}
\xi_{v}\left(\mathbf{x}\right)=\sum_{c}A_{vc}\psi_{c}\left(\mathbf{x}\right)
\end{equation}
where $\mathbf{x}$ is a combined spin and position coordinate. In
this way we can represent a generic excited state with $N_{v}$ $\xi_{v}$
functions which are represented in the code in the same way as wavefunctions
with just the restriction of belonging to the manifold of empty KS
states.

Instead of working with the KS valence states it can be convenient
to work with their representation as maximally localised Wannier's
functions $\left\{ w_{v}\right\} $\cite{Marzari97}:

\begin{equation}
w_{v}=\sum_{v'}U_{vv'}\psi_{v'}
\end{equation}
where $U$ is an unitary matrix:

\begin{equation}
\psi_{v}=\sum_{v'}U_{vv'}^{-1}w_{v'}=\sum_{v'}U_{v'v}^{*}w_{v'}
\end{equation}
We can now express the excitonic state as:

\begin{equation}
\left|\Theta\right\rangle =\sum_{v}\left[\hat{\tilde{a}}_{v}\sum_{c}\left(\left\langle \psi_{c}\right|\tilde{\xi_{v}}\left.\right\rangle \hat{a}_{c}^{\dagger}\right)\right]\left|\Phi_{0}\right\rangle 
\end{equation}
with the definitions:
\begin{equation}
\hat{\tilde{a}}_{v}=\left[\int d\mathbf{x}\hat{\Psi}\left(\mathbf{x}\right)w_{v}\left(\mathbf{x}\right)\right]=\sum_{v'}U_{vv'}\hat{a}_{v}
\end{equation}
and

\begin{equation}
\tilde{\xi}_{v}=\sum_{v'}U_{vv'}^{*}\xi_{v'}
\end{equation}
In the following, we will focus on the general case of time-reversal
symmetric systems. In this case the one-body states $\psi_{i}$ and
the $U$ matrix can be taken real and we will omit in the following
complex-conjugate signs. It is worth noting how the introduced transformation
for the $\xi_{v}$'s works on the spatial representation $\Theta\left(\mathbf{x},\mathbf{x}'\right)$
of the excitonic state:

\begin{equation}
\left|\Theta\right\rangle =\int d\mathrm{x}d\mathrm{x}'\hat{\Psi}\left(\mathbf{x}\right)\hat{\Psi}^{\dagger}\left(\mathbf{x}'\right)\Theta\left(\mathbf{x},\mathbf{x}'\right)
\end{equation}
with:

\begin{equation}
\Theta\left(\mathbf{x},\mathbf{x}'\right)=\sum_{vc}\psi_{v}\left(\mathbf{x}\right)\sum_{c}A_{vc}\psi_{c}\left(\mathbf{x}'\right)=\sum_{v}\psi_{v}\left(\mathbf{x}\right)\xi_{v}\left(\mathbf{x}'\right)
\end{equation}
it results:

\begin{equation}
\Theta\left(\mathbf{x},\mathbf{x}'\right)=\sum_{v}w_{v}\left(\mathbf{x}\right)\tilde{\xi}_{v}\left(\mathbf{x}'\right)
\end{equation}

We will now focus on the common case in which spin-orbit coupling
can be neglected, we can separate spatial and spin coordinates and
the ground state $\left|\Phi_{0}\right\rangle $ is built from $N_{v}$
doubly occupied KS states. In this case the eigen-states of $\hat{H}^{eh}$
will be either spin-singlet or spin-triplet states. In this case it
is convenient to retain in the description of the excitonic state
only the spatial coordinates. A generic excitonic states will be described
by a set of $N_{v}$ $\left\{ \xi_{v}\left(\mathbf{r}\right)\right\} $
functions. The $\hat{H}^{eh}$ operator acting on these states involves
only spatial coordinates but assumes a different form for spin-singlet
and spin-triplet states:

\begin{equation}
\hat{H}_{singlet}^{eh}=\hat{D}+2\hat{K}^{x}+\hat{K}^{d}
\end{equation}

\begin{equation}
\hat{H}_{triplet}^{eh}=\hat{D}+\hat{K}^{d}
\end{equation}
where $\hat{D}$ is referred to as the diagonal operator, $\hat{K}^{x}$
as the exchange one and $\hat{K}^{d}$ as the direct one.

We show now how these operators act on the $\left\{ \xi_{v}\left(\mathbf{r}\right)\right\} $
and $\left\{ \tilde{\xi}_{v}\left(\mathbf{r}\right)\right\} $ representations
of the excitonic state which we indicate using the bra-ket form as
$\left|\left\{ \xi\right\} \right\rangle $ and $\left|\left\{ \tilde{\xi}\right\} \right\rangle $\cite{Rocca10}.
We start from the diagonal term $\hat{D}$ in the $A_{vc}$ representation
:
\begin{equation}
A'_{vc}=\sum_{v'c'}D_{vc,v'c'}A_{v'c'}
\end{equation}
with:

\begin{equation}
D_{vc,v'c'}=\left(\epsilon_{c}-\epsilon_{v}\right)\delta_{v,v'}\delta_{c,c'}
\end{equation}
and we can write:

\begin{equation}
\left|\left\{ \xi'\right\} \right\rangle =\hat{D}\left|\left\{ \xi\right\} \right\rangle 
\end{equation}
with 
\begin{equation}
\left|\xi_{v}'\right\rangle =\left(\hat{H}^{q.p.}-\epsilon_{v}\right)\left|\xi_{v}\right\rangle 
\end{equation}
where we suppose that the quasi-particle energies $\epsilon_{i}$
are given by the eigen-values of the Hamiltonian operator $\hat{H}^{q.p.}$
whose eigenfunctions coincide with the KS states as we are using the
diagonal GW approximation. Usually such an operator can be obtained
from the Kohn-Sham Hamiltonian $H_{KS}$ adding one or more scissor
terms. It is worth noting that the $\left\{ \xi_{v}'\right\} $ functions
remain confined on the manifold of unoccupied KS states. The computational
cost of applying the $\hat{D}$ operator to a generic state $\left|\left\{ \xi\right\} \right\rangle $
scales as $N^{2}\log N$ with respect to the generic system size $N$.
Indeed, applying the $\hat{H}^{q.p.}$ operator to a single $\xi_{v}$
function has the same $N\log N$ scaling of applying a KS Hamiltonian
while the number $N_{v}$ of valence states scales as $N$.

For the $\hat{K}^{x}$ operator acting on the $\left|\left\{ \xi\right\} \right\rangle $
representation of an excitonic state, we start from the $A_{vc}$
representation:

\begin{equation}
K_{vc,v'c'}^{x}=\int\psi_{v}\left(\mathbf{r}\right)\psi_{c}\left(\mathbf{r}\right)v\left(\mathbf{r},\mathbf{r}'\right)\psi_{v'}\left(\mathbf{r}'\right)\psi_{c'}\left(\mathbf{r}'\right)d\mathbf{r}d\mathbf{r}'
\end{equation}
where $v$ is the bare Coulomb operator (we recall that we consider
real wavefunctions). We can write for :

\begin{equation}
\left|\left\{ \xi'\right\} \right\rangle =\hat{H}^{x}\left|\left\{ \xi\right\} \right\rangle 
\end{equation}

\begin{equation}
\xi_{v}'\left(\mathbf{r}\right)=\int P_{c}\left(\mathbf{r},\mathbf{r}'\right)\psi_{v}\left(\mathbf{r'}\right)v\left(\mathbf{r}',\mathbf{r}''\right)\psi_{v'}\left(\mathbf{r}''\right)\xi_{v'}\left(\mathbf{r}''\right)d\mathbf{r'}d\mathbf{r}''
\end{equation}
where we used the projector operator on the manifold of unoccupied
one-body states:

\begin{equation}
P\left(\mathbf{r},\mathbf{r}'\right)=\sum_{c}\psi_{c}\left(\mathbf{r}\right)\psi_{c}\left(\mathbf{r}'\right)=\delta\left(\mathbf{r}-\mathbf{r}'\right)-\sum_{v}\psi_{v}\left(\mathbf{r}\right)\psi_{v}\left(\mathbf{r}'\right)\label{eq:hx1}
\end{equation}
Hence, the evaluation of the $\left\{ \xi_{v}'\right\} $ in Eq. \ref{eq:hx1}
does not require the calculation of any unoccupied one-body state.
The computational cost of applying the $\hat{H}^{x}$ operator to
a generic state $\left|\left\{ \xi\right\} \right\rangle $ scales
as $N^{3}$. Indeed, the evaluation of the function:
\begin{equation}
\rho'\left(\mathbf{r}\right)=\sum_{v}\psi_{v}\left(\mathbf{r}\right)\xi_{v}\left(\mathbf{r}\right)
\end{equation}
scales as $N^{2}$ and applying the $\hat{P}_{c}$ projector to a
single function scales as $N^{2}$ .

We show now how the $\hat{K}^{d}$ operator acts on the$\left|\left\{ \xi\right\} \right\rangle $
representation of an excitonic state. We start from the $A_{vc}$
representation :

\begin{equation}
K_{vc,v'c'}^{d}=-\int\psi_{c}\left(\mathbf{r}\right)\psi_{c'}\left(\mathbf{r}\right)W\left(\mathbf{r},\mathbf{r}'\right)\psi_{v'}\left(\mathbf{r}'\right)\psi_{v}\left(\mathbf{r}'\right)d\mathbf{r}d\mathbf{r}'
\end{equation}
where $W$ is the static screened Coulomb interaction, and we write:

\begin{equation}
\left|\left\{ \xi'\right\} \right\rangle =\hat{H}^{d}\left|\left\{ \xi\right\} \right\rangle 
\end{equation}
\begin{eqnarray*}
\xi_{v}'\left(\mathbf{r}\right) & = & -\int P_{c}\left(\mathbf{r},\mathbf{r}'\right)\sum_{v'}\xi_{v'}\left(\mathbf{r}'\right)\times\\
 &  & \left[\int W\left(\mathbf{r'},\mathbf{r}''\right)\psi_{v'}\left(\mathbf{r}''\right)\psi_{v}\left(\mathbf{r}''\right)d\mathbf{r}''\right]d\mathbf{r}'
\end{eqnarray*}
with an overall computational cost scaling as $N^{4}$ due to the
term in square brackets. In order to reduce the computational cost
it is advantageous to turn to the Wannier's representation of the
excitonic states:
\begin{equation}
\left|\left\{ \tilde{\xi}'\right\} \right\rangle =\hat{H}^{d}\left|\left\{ \tilde{\xi}\right\} \right\rangle 
\end{equation}
\begin{eqnarray*}
\tilde{\xi}_{v}'\left(\mathbf{r}\right) & = & -\int P_{c}\left(\mathbf{r},\mathbf{r}'\right)\sum_{v'}\tilde{\xi}_{v'}\left(\mathbf{r}'\right)\times\\
 &  & \left[\int W\left(\mathbf{r'},\mathbf{r}''\right)w_{v'}\left(\mathbf{r}''\right)w_{v}\left(\mathbf{r}''\right)d\mathbf{r}''\right]d\mathbf{r}'
\end{eqnarray*}
the computational cost can now be lowered to $N^{3}$ scaling if in
the term in square brackets only terms involving overlapping couples
of Wannier's functions are evaluated. Indeed, we neglect in the evaluation
of $\hat{K}^{d}$ the calculation of terms involving products of non-overlapping
Wannier's functions and we indicate with $s$ the threshold which
determines whether two Wannier's function $w_{v}$ and $w_{v'}$ overlap:

\begin{equation}
\int\left|w_{v}\left(\mathbf{r}\right)\right|^{2}\left|w_{v'}\left(\mathbf{r}\right)\right|^{2}d\mathbf{r}>s\label{eq:overlap}
\end{equation}
The $\hat{K}^{d}$ operator is usually separated into a bare $K^{d,b}$
and a correlation part $K^{d,c}$ according to the separation of the
$W$ operator as:

\begin{equation}
W\left(\mathbf{r},\mathbf{r}'\right)=v\left(\mathbf{r},\mathbf{r}'\right)+W_{c}\left(\mathbf{r},\mathbf{r}'\right)
\end{equation}
where $W_{c}$ is the correlation part of the screened Coulomb interaction.
Let's inspect the first term:

\begin{equation}
\left|\left\{ \tilde{\xi}'\right\} \right\rangle =\hat{H}^{d,b}\left|\left\{ \tilde{\xi}\right\} \right\rangle 
\end{equation}

\begin{equation}
\tilde{\xi}'_{v}\left(\mathbf{r}\right)=-\int P_{c}\left(\mathbf{r},\mathbf{r}'\right)\sum_{v'}\tau_{vv'}^{b}\left(\mathbf{r}'\right)\tilde{\xi}_{v'}\left(\mathbf{r}'\right)d\mathbf{r}'
\end{equation}
with:

\begin{equation}
\tau_{vv'}^{b}\left(\mathbf{r}\right)=\int v\left(\mathbf{r},\mathbf{r}'\right)w_{v'}\left(\mathbf{r}'\right)w_{v}\left(\mathbf{r}'\right)d\mathbf{r}'
\end{equation}
which is evaluated only for overlapping couples $v,v'$ of Wannier's
functions. Let's inspect the second term:
\begin{equation}
\left|\left\{ \tilde{\xi}'\right\} \right\rangle =\hat{H}^{d,c}\left|\left\{ \tilde{\xi}\right\} \right\rangle 
\end{equation}

\begin{equation}
\tilde{\xi}'_{v}\left(\mathbf{r}\right)=-\int P_{c}\left(\mathbf{r},\mathbf{r}'\right)\tau_{vv'}^{c}\left(\mathbf{r}'\right)\tilde{\xi}_{v'}\left(\mathbf{r}'\right)d\mathbf{r}'
\end{equation}
with:

\begin{equation}
\tau_{vv'}^{c}\left(\mathbf{r}\right)=\int W_{c}\left(\mathbf{r},\mathbf{r}'\right)w_{v'}\left(\mathbf{r}'\right)w_{v}\left(\mathbf{r}'\right)d\mathbf{r}'
\end{equation}
which is evaluated only for overlapping couples $v,v'$ of Wannier's
functions. In our present implementation, we express $W_{c}$ starting
from the screened polarisability operator $\Pi$, in the static approximation:

\begin{equation}
W_{c}\left(\mathbf{r},\mathbf{r}'\right)=\int v\left(\mathbf{r},\mathbf{r}''\right)\Pi\left(\mathbf{r''},\mathbf{r}'''\right)v\left(\mathbf{r}''',\mathbf{r}'\right)d\mathbf{r}''d\mathbf{r}'''
\end{equation}
and we use an \emph{optimal basis set} $\left\{ \Phi_{\mu}\right\} $
for representing $\Pi$\cite{Umari09,Umari10}:

\begin{equation}
\Pi\left(\mathbf{r},\mathbf{r}'\right)=\sum_{\mu\nu}\Phi_{\mu}\left(\mathbf{r}\right)\Pi_{\mu\nu}\Phi_{\nu}\left(\mathbf{r}'\right)
\end{equation}
We can now write:

\begin{equation}
W_{c}\left(\mathbf{r},\mathbf{r}'\right)=\sum_{\mu\nu}\left(v\Phi_{\mu}\right)\left(\mathbf{r}\right)\Pi_{\mu\nu}\left(v\Phi_{\nu}\right)\left(\mathbf{r}'\right)
\end{equation}
with the notation:
\begin{equation}
\left(v\Phi_{\mu}\right)\left(\mathbf{r}\right)=\int d\mathbf{r}'v\left(\mathbf{r},\mathbf{r}'\right)\Phi_{\mu}\left(\mathbf{r}'\right)
\end{equation}
We express the $\tau_{vv'}^{c}$ terms as:
\begin{equation}
\tau_{vv'}^{c}\left(\mathbf{r}\right)=\sum_{\mu}\left(v\phi_{\mu}\right)\left(\mathbf{r}\right)Z_{v'\mu}^{v}
\end{equation}
with the definition:

\begin{equation}
Z_{v'\mu}^{v}=\sum_{\nu}\Pi_{\mu\nu}\int d\mathbf{r}\left(v\phi_{\nu}\right)\left(\mathbf{r}\right)w_{v}\left(\mathbf{r}\right)w_{v'}\left(\mathbf{r}\right)
\end{equation}

We have implemented our method as a module of the Quantum-Espresso
(Q-E) DFT package\cite{Q-E} which is based on the plane-waves pseudopotentials
paradigm. We sample the Brillouin's zone at the sole $\Gamma$ point
and we can treat both extended (through periodic boundary conditions)
and isolated systems. Quasi-particle energies are taken from a GW
calculation for a set of conduction and valence states while for the
residual states we apply to the DFT ones a scissor operator \cite{Rocca10}.
The static screened Coulomb interaction is provided by a GW calculation
performed with the GWL code which overcomes the problem of sums over
empty states\cite{Umari09,Umari10}. Maximally localised Wannier's
function are obtained using the algorithm described in Ref. \cite{Gygi03}.
Single excitonic states can be calculated through a conjugate gradient
minimization while the complex dielectric function can be directly
evaluated through a Lanczos series\cite{Ankudinov02}.

\section{Validation}

First, we validated our method considering the isolated benzene molecule
comparing the first excitonic energies with those calculated with
the Yambo code \cite{Yambo} starting from the same Q-E DFT. We find
that values from the two calculations agree within 0.1 eV.\textcolor{black}{{}
As our implementation, but the O($N^{3}$) scaling due to the Wannier's
representation of excitonic amplitudes, is equivalent to ordinary
plane-waves pseudo-potential BSE codes, the same good quality of results
is expected for generic systems.}

We then applied our BSE scheme to address the isolated C$_{60}$ molecule.
We take the atomic coordinates from Ref. \cite{Qian15} and choose
the PBE exchange and correlation functional \cite{Perdew:1996} for
the starting DFT calculation together with a norm-conserving pseudo-potential
\cite{C60pseudo}. The cubic simulation cell has an edge of 40 Bohr
and the Coulomb interaction is truncated at $20$ Bohr. In Ref. \cite{Qian15}
we investigated the GW quasi-particle energy levels finding a self-consistent
HOMO-LUMO (electronic) gap of 4.94 eV (1.78 eV at the PBE level) and
we consider the corresponding scissor operator in the BSE calculation.
A basis of 2000 elements is used for representing the static polarisability
operator \cite{Umari09,Umari10}.

\textcolor{black}{For evaluating the C$_{60}$ optical band gap, we
obtained the lowest in energy optically allowed singlet exciton energy
and the corresponding amplitude through a conjugate gradient minimisation.
While, for evaluating optical absorption spectra, we obtained the
complex dielectric susceptibility function through the Lanczos algorithm.
}As C$_{60}$ exhibits 120 doubly occupied valence states the maximum
number of product terms is 14400. In Tab. \ref{tab:Optical-band-gap C60},
we report the calculated values for the optical gap as a function
of the $s$ threshold.

\begin{figure}
\includegraphics[scale=0.18]{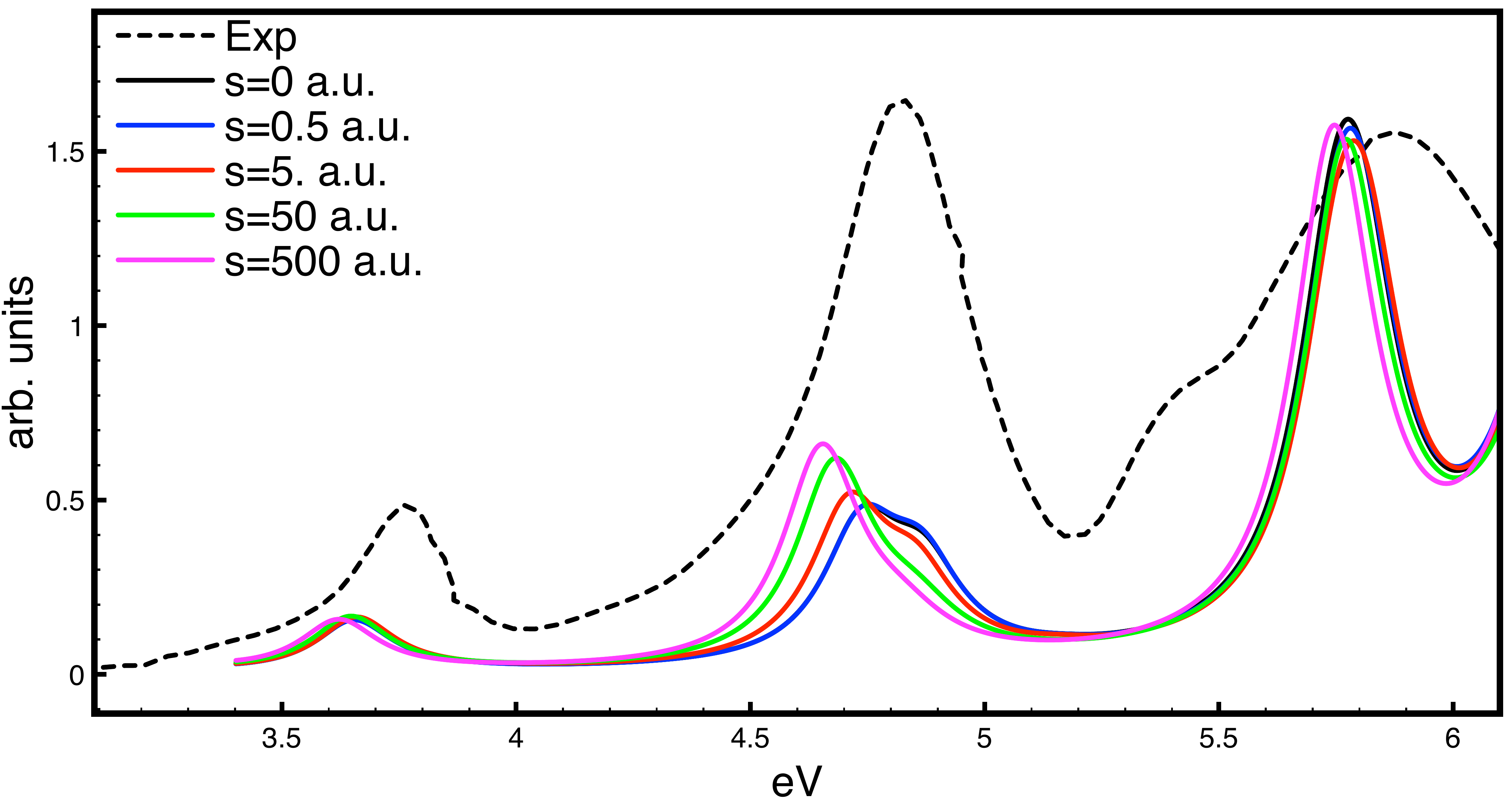}

\protect\caption{\label{fig:Optical-absorption-spectrum}Optical absorption spectrum
of molecular C$_{60}$: Experiment from Ref. \cite{Bauernschmitt98}(black-dashed);
GW-BSE (solid) with:$s=0$ a.u. (black), $s=0.5$ a.u. (blue),$s=5$
a.u. (red), $s=50$ a.u. (green), $s=500$ a.u. (magenta). A Lorentzian
broadening of 0.1 eV has been applied.}
\end{figure}

\begin{table}
\begin{tabular}{|c|c|c|}
\hline 
$s$ (a.u.) & \#Products & $E_{g}$ (eV)\tabularnewline
\hline 
\hline 
2500 & 90 & 3.67\tabularnewline
\hline 
500 & 180 & 1.86\tabularnewline
\hline 
50 & 900 & 1.87\tabularnewline
\hline 
5 & 2820 & 1.90\tabularnewline
\hline 
0.5 & 6816 & 1.92\tabularnewline
\hline 
0.05 & 11524 & 1.92\tabularnewline
\hline 
0. & 14400 & 1.92\tabularnewline
\hline 
\end{tabular}

\protect\caption{Optical band\label{tab:Optical-band-gap C60} gap for C$_{60}$ for
various $s$ thresholds with corresponding numer of Wannier's functions
products. }
\end{table}
We observed that a large threshold $s$ of 500 a.u., corresponding
to only 180 product terms, was sufficient for converging the optical
gap $E_{g}$ within 0.06 eV with respect to the final value of 1.92
eV. This is in excellent agreement with the experimental figures of
1.92 eV (solution) and of 1.87 eV (bulk solid) \cite{Wang95} and
with a previous BSE value of 1.88 eV\cite{Tiago08} . The same fast
convergence with $s$ is observed for the optical absorption spectrum
reported in Fig. \ref{fig:Optical-absorption-spectrum}. Also in this
case a threshold $s=500$ a.u. yields results close (within 0.1 eV
for peak positions) to the fully-converged  $s=0$ a.u. calculation
involving all possible product terms. We notice also good agreement
with  experimental data in solution \cite{Bauernschmitt98} for the
position and relative intensities of the main features.

\section{Application to JK2@TiO$_{2}$}

\begin{table*}[t]
\begin{tabular}{|c|c|c|c|c|c|c|}
\hline 
$s$ (a.u.) & \#Products & $E_{g}$ (eV) & Tot. Time (s) & Time $K^{x}$ (s) & Time $K^{d}$ (s) & Single $K^{d}$ (s)\tabularnewline
\hline 
\hline 
500 & 2295 & 2.40 & 254580 & 43902 & 53079 & 23.1\tabularnewline
\hline 
50 & 7653 & 2.43 & 316082 & 43677 & 116256 & 15.2\tabularnewline
\hline 
10 & 12711 & 2.47 & 411780 & 44024 & 208490 & 16.4\tabularnewline
\hline 
\end{tabular}

\protect\caption{\label{tab:GW-BSE-calculations-of}GW-BSE calculations of the JK2@TiO$_{2}$
system for different values of the threshold $s$: number of Wannier's
functions products (\#Products), optical band gap ($E_{g}$); calculation
times \cite{Computer}: total (Tot. time), for the exchange term $K^{x}$
(Time $K^{x}$ ), for the direct term $K^{d}$ (Time $K^{d}$ ), ratio
Time $K^{d}$ / \#Products (Single $K^{d}$ )}
\end{table*}

\begin{figure}
\includegraphics[scale=0.18]{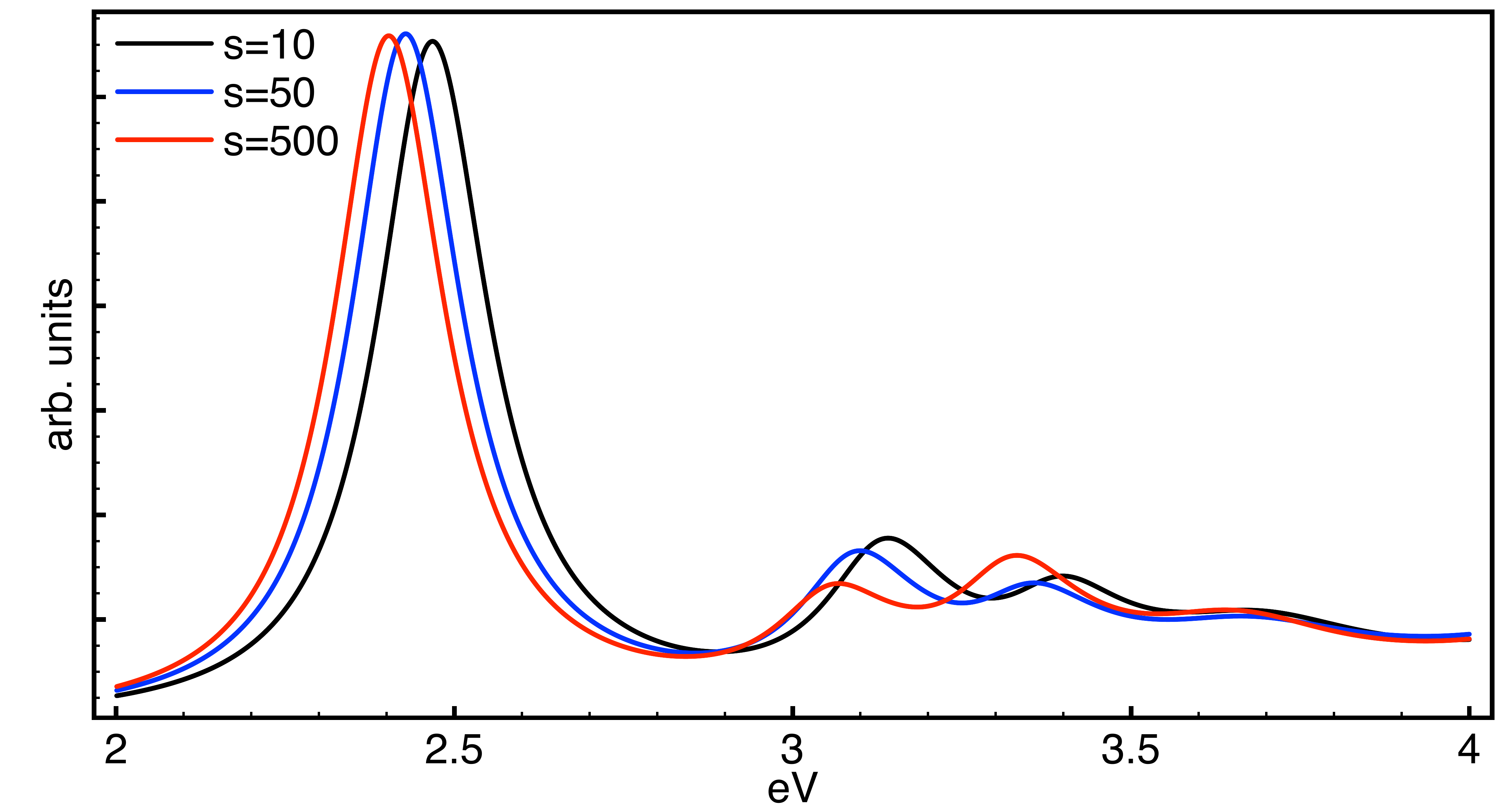}

\protect\caption{\label{fig:Optical-absorption-spectrum-1}Optical absorption spectrum
for the JK2@TiO$_{2}$ model:GW-BSE with:$s=10$ a.u. (black), $s=50$
a.u. (blue), , $s=500$ a.u. (red). A Lorentzian broadening of 0.1
eV has been applied.}

\end{figure}

\begin{table}
\begin{tabular}{|l|l|l|}
\hline 
 & Exp (eV)  & \multicolumn{1}{c|}{GW-BSE (eV)}\tabularnewline
\hline 
\hline 
D/D$^{+}$ (eV) & -5.4 & -5.9 (-5.8)\tabularnewline
\hline 
CBM & -4.0 & -4.3 (-4.1)\tabularnewline
\hline 
$E_{gap}$ (eV) & 2.7 & 2.5 (2.5)\tabularnewline
\hline 
\end{tabular}

\protect\caption{\label{tab:Energy-levels-for}Energy levels for the JK2@TiO$_{2}$
system: D/D$^{+}$ ground state dye ionisation potential, CBM TiO$_{2}$
conduction band minimum, $E_{gap}$ optical gap. Experimental figure
are from Ref. \cite{Kim06}. GW-BSE figures in parentheses: including
solvent effects. }
\end{table}
We can now apply our GW-BSE approach for evaluating the key energy
levels in a DSSC model. We considered the same JK2@TiO$_{2}$ model
structure of Ref. \cite{Pastore13} exhibiting a dye molecule adsorbed
on an anatase TiO$_{2}$ cluster (32 TiO$_{2}$ units) comprising
204 atoms and 1174 valence electrons in a $35\times35\times66.5$
Bohr$^{3}$ periodic orthorhombic simulation cell. DFT calculations
are performed using the BLYP exchange and correlation functional \cite{BLYP}
and wavefunctions are expanded on a plane waves basis set defined
by a cutoff energy of 150 Ry. For the GW and the BSE calculations
the polarisability operators are expanded on a (minimal) optimal basis
set \cite{Umari10} comprising 2500 vectors corresponding to an estimated
accuracy for the GW energy levels of $\sim0.2$ eV \cite{estar}.
The BSE calculation has been performed using thresholds down to $s=10$
a.u. The optical gap $E_{gap}$ is determined through the position
of the first peak in the calculated optical absorption spectrum consistently
with the reported experimental data. In Tab. \ref{tab:GW-BSE-calculations-of}
we show that a threshold $s=500$ a.u. assures convergence within
$\sim0.1$ eV as in the case of C$_{60}$ involving only 2295 product
terms to be contrasted to the total number of possible product terms
$562^{2}=315844$. The same degree of accuracy is found for the absorption
spectrum as displayed in Fig.\ref{fig:Optical-absorption-spectrum-1}.

In analogy with Refs. \cite{Umari13,Verdi14}, the dye D/D$^{+}$
level (neglecting any structural relaxation) is given by the uppermost
occupied GW energy level with dye character which corresponds also
to the HOMO of the entire JK2@TiO$_{2}$ system. The D$^{*}$/D$^{*+}$
level is the excited state ionisation potential of the adsorbed dye
which can be obtained from the D/D$^{+}$ level adding the BSE lowest
bright (singlet) excitation energy of the adsorbed dye. The evaluation
of the TiO$_{2}$ CBM is more delicate. Indeed, the electronic VBM-CBM
gap converges very slowly towards the bulk value when calculated on
slabs or clusters of increasing size. Moreover, this issue is more
severe for the GW approach \cite{Freysoldt08}. However, we previously
found that the absolute position of the VBM in anatase TiO$_{2}$
slabs converges faster with respect to slab width and is in agreement
with available experimental data \cite{VBM}. This is related to the
localisation of the manifold of occupied states. Hence, a good estimate
for the TiO$_{2}$ CBM can be obtained by adding to the VBM value,
calculated for the cluster, the electronic band gap of bulk anatase
TiO$_{2}$ from an equivalent GW calculation for a bulk model structure
(3.6 eV \cite{Umari13}).

As displayed in Tab. \ref{tab:Energy-levels-for} the GW-BSE energy
levels are in a quite good agreement with experiment even without
considering any solvent effect. These can be estimated considering
the shift of the energy levels at the sole DFT BLYP level using an
implicit solvent model. This further increases the quality of the
GW-BSE results showing that the GW-BSE approach can predict at the
same time both charged and neutral energy levels.\textcolor{black}{{}
We note that our GW-BSE results improve on the best TD-DFT ones presented
in Ref. \cite{Pastore13} for the same JK2@TiO$_{2}$ system which
strongly depend on the flavour of the adopted hybrid exchange and
correlation functional. We don\textquoteright t have such arbitrariness
in BSE as such method as been shown to give good results both for
isolated and extended systems in contrast to what is recorded for
TD-DFT. }

\section{Conclusions}

It is worth stressing that with our method a BSE calculation performed
on a large system as JK2@TiO$_{2}$ was made possible even with quite
restricted computational resources \cite{Computer}. Indeed, for this
system the total number of possible Wannier's functions products is
huge (344569). In Fig. \ref{fig:Optical-absorption-spectrum-1} and
Tab. \ref{tab:GW-BSE-calculations-of}, we show that only a fraction
of these products is required for assuring good convergence while
the cost for evaluating the $\hat{K}^{d}$ term of the excitonic Hamiltonian
scales almost linearly with such number\cite{cache}. Indeed, a threshold
$s=500$ a.u., corresponding to $\sim4$ product terms for each Wannier's
function, leads to pretty well converged results. Considering all
the possible Wannier's functions products would require a $\sim150$
times longer computation together with much more severe requirements
in terms of memory usage.

\textcolor{black}{We note that a constant number of product terms
for each Wannier's function determines a theoretical O(N$^{3}$) scaling
based on the computational cost of the terms appearing in the excitonic
Hamiltonian. We checked that this relation is observed when comparing
two analogous C$_{60}$ and JK2@TiO$_{2}$ BSE Lanzos (200 steps)
calculations for the same $s=500$ a.u. threshold which gives comparable
accuracy. The theoretically expected cost of the JK2@TiO$_{2}$ ($T_{JK2}^{*}$)
calculation can be obtained from the computing time of the C$_{60}$
one ($T_{C_{60}}$) :}

\textcolor{black}{
\[
T_{JK2}^{*}=\left(\frac{N_{\mathrm{cores}}^{JK2}}{N_{\mathrm{cores}}^{C_{c0}}}\right)\left(\frac{N_{PW}^{JK2}}{N_{PW}^{C_{60}}}\right)\left(\frac{N_{v}^{JK2}}{N_{v}^{C_{60}}}\right)^{3}T_{C_{60}},
\]
where $N_{\mathrm{core}}$ is the number of used computing cores,
$N_{v}$ the number of valence states, and $N_{PW}$ the ratio between
number of plane waves and of valence states. The $(N_{PW}^{JK2}/N_{PW}^{C60})$
ratio must be added because of the different energy cutoffs used in
C$_{60}$ and in JK2@TiO$_{2}$ for defining the plane waves basis
sets: 45 and 150 Ry, respectively. The relevant figures for C$_{60}$
are: $N_{\mathrm{cores}}^{C_{60}}=32$, $N_{PW}^{C60}=2716$, and
$T_{C_{60}}=2070$ s. The corresponding figures for JK2@TiO$_{2}$
are: $N_{\mathrm{cores}}^{JK2}=64$, $N_{PW}^{JK2}=4403$, and $T_{JK2}=254580$
s. Hence, the estimated computing time found for JK2@TiO$_{2}$ is
$T_{JK2}^{*}=196395$ s, in excellent agreement with the recorded
one (254580 s).}

In conclusion, we presented a method which permits to extend the scope
of state of the art BSE calculations even using limited computational
resources.\textcolor{black}{{} This is achieved by applying the unitary
transformation, which defines maximally localised Wannier's functions
of valence KS orbitals, to excitonic amplitudes. }The accuracy is
controlled by a single parameter with a clear physical meaning (i.e.
defining the overlap of Wannier's functions). Moreover, its formulation
is general and can be extended towards other applications as TD-DFT.
Further speed-up could be obtained restricting integrals in real space
only to the volumes where Wannier's functions are not vanishing.

This permitted us to solve the problem of energy level alignment in
JK2@TiO$_{2}$ opening the way to the use of BSE for addressing complex
systems in the modelling of photovoltaics and optoelectronic materials
where ordinary approaches fail.


\begin{thebibliography}{10}
\bibitem{C60pseudo}We use the C.pbe-tm.UPF pseudo-potential from
the Q-E repository.
\bibitem{Computer}Calculations were run on an INTEL Xeon 2.6 GHz
64 cores computer cluster with 8GB RAM per core.
\bibitem{cache}Computing $K^{d}$ for a single Wannier's functions
product is slightly longer for small and large $s$ for the worse
use of cache and swap memory, respectively.
\bibitem{VBM}We repeated the calculations of Ref. \cite{Umari13}
with a dye@TiO$_{2}$ model comprising 3 TiO$_{2}$ layers instead
of 2. The VBM changes of only +0.1 eV.
\end{thebibliography}
\end{document}